\documentclass[a4paper,11pt]{article}
\usepackage{pos}

\title{Theory of Gamma-Ray Loud AGNs}

\author[a,b]{Frank M. Rieger}

\affiliation[a]{Institute for Theoretical Physics, Heidelberg University,
Philosophenweg 12, 69120 Heidelberg, Germany}
\affiliation[b]{Max-Planck-Institute for Nuclear Physics (MPIK), 
P.O. Box 103980, 69029 Heidelberg, Germany}

\emailAdd{f.rieger@uni-heidelberg.de}

\abstract{The last decade has seen tremendous developments in gamma-ray astronomy 
with the extragalactic sky becoming highly populated by Active Galactic Nuclei (AGNs). 
This brief review highlights some of the progress in AGN research achieved over the 
years, and discusses exemplary advances in the theory and physics of gamma-ray loud 
AGNs, including black-hole magnetospheric processes, the physics of pc-scales jets, as 
well as particle acceleration and high-energy emission in the large-scale jets of AGNs.}

\FullConference{%
  7th Heidelberg International Symposium on High-Energy Gamma-Ray Astronomy 
  (Gamma2022)\\
  4-8 July 2022\\
  Barcelona, Spain\\}

\newcommand{\lppr}{\stackrel{<}{\scriptstyle \sim}}

\begin{document}
\maketitle

\section{Introduction}
Radio-loud or {\it jetted AGNs} \cite{Padovani2017}, i.e. AGNs with strong relativistic jets, are 
the most persistent, powerful sources in the Universe. Their detection at gamma-ray energies 
reveals that a significant amount of their power is deposited a highest energies, and demonstrates 
them to be exceptional cosmic particle accelerators \cite{Biteau2020}. All types of jetted AGNs 
have been seen at gamma-ray energies, from more misaligned radio galaxies to blazar-type 
sources where the jets are seen face-on \cite{Urry1995,Costamante2020,Rulten2022}.\\
Gamma-ray astrophysics is thus expected to play a fundamental role in resolving key issues 
in AGN physics, such as: {\it (i) How are relativistic jets being formed?} Are they preferentially
ergospheric or disk-driven? If the former, what is their plasma source; if the latter, what is their 
accretion-disk connection? 
{\it (ii) What makes jets radiate?} What is the dominant gamma-ray radiation process and 
particle acceleration mechanism? Where is it located? What is the plasma composition?
{\it (iii) How are small and large scales connected?} How is the energy transported from
the black hole (BH) to the outer lobes scales? How are jets confined? What kind of 
instabilities are important?

\section{On the Extragalactic Gamma-Ray Sky}
\begin{figure}[htb]
\begin{center}
\includegraphics[width = 0.7 \textwidth]{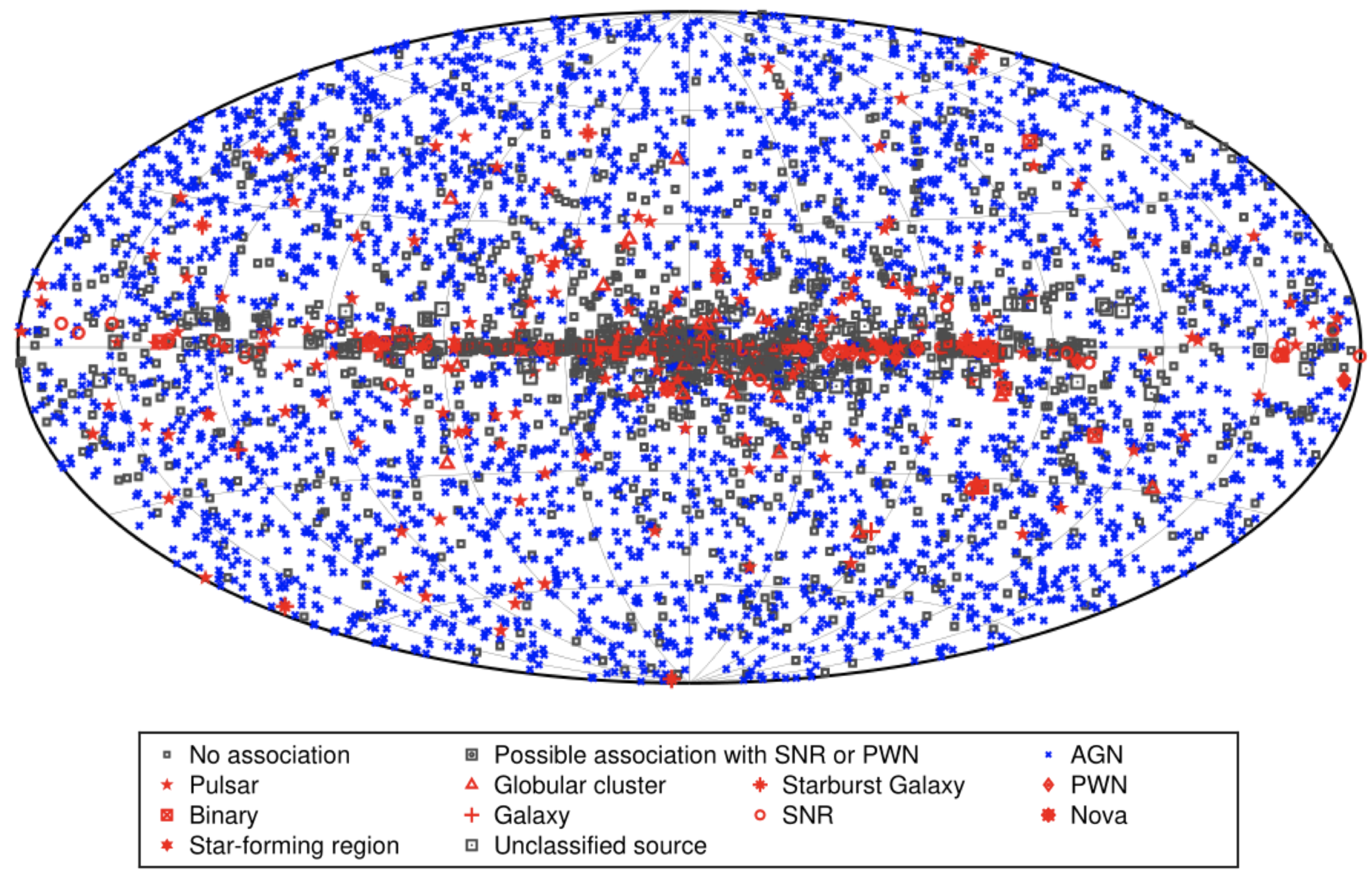}
\caption{High-energy sky map (4FGL) based on 8 yr of Fermi-LAT data. All AGNs (all 
classes) are plotted with the same blue symbol. From ref.~\cite{4FGL}. \copyright AAS. 
Reproduced with permission.}
\label{4FGL}
\vspace*{-0.5cm}
\end{center}
\end{figure}

\begin{figure}[htb]
\begin{center}
\includegraphics[width = 1.0 \textwidth]{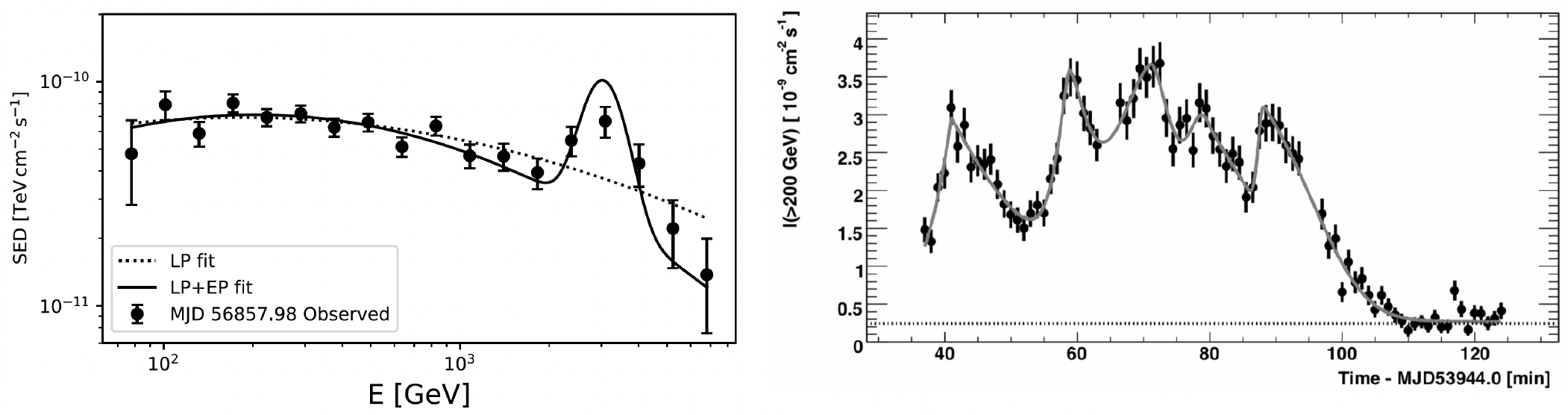}
\caption{{\it Left:} The VHE SED of Mkn 501 ($z=0.034$) during an elevated state in 2014 
July 19 as measured by MAGIC, revealing an unusual feature at $\sim3$ TeV. From 
ref.~\cite{MAGIC_MKN501}, reproduced with permission~\copyright~ESO. {\it Right:} The 
VHE light curve of PKS~2155-304 ($z=0.116$) during its famous outburst in July 2006 as 
seen by H.E.S.S., revealing substantial flux changes down to $\sim 3$ min. From 
ref.~\cite{HESS_PKS2155}.}
\label{capabilities}
\vspace*{-0.5cm}
\end{center}
\end{figure}

The extragalactic sky has become bright at gamma-ray energies. The Fermi-LAT 12 year 
point source catalog (4FGL-DR3) for example, reports the detection of 6658 sources at 
high energies (HE) $>$ 50 MeV, with more than 3740 identified as belonging to the blazar 
class of AGNs, cf. Fig.~\ref{4FGL}.
About 70 sources have been identified as non-blazar AGNs, out of which 45 are radio galaxies 
(among them 22 Fanaroff-Riley (FR)~I and 14 FR~II) and 8 are Narrow Line Seyfert 1 (NLSy~1) 
galaxies \cite{4FGL,4FGL-3DR}. At very high energies (VHE; $\geq 100$ GeV), about 85 AGNs 
are listed in the TeVCat catalog, with redshifts up to $z\simeq 1$. While most of the TeV sources 
are BL Lac objects (55 HBL, 10 IBL, 2 LBL), a few prominent radio galaxies (e.g., Cen A, M87 
and NGC 1275) are present as well. The latter sources have raised considerable interest by
allowing insights into the immediate vicinity of the supermassive BH environment in AGNs 
(cf. Sec.~4.1 below). 
For a recent review of the physics case of gamma-ray emitting, non-blazar AGNs, the reader 
is referred to ref.~\cite{RiegerL2018}.\\
Spectral and timing capabilities at gamma-ray energies have significantly developed over time, 
allowing, in some sources, to probe rapid variability down to a few minutes and unusual spectral 
features that could signal new physical processes, see e.g., Fig.~\ref{capabilities}. 
                                                  
The detection of NLSy~1 galaxies at HE energies has raised several issues.
Radio-quiet (non-jetted) NLSy~1 are commonly thought to be high-Eddington 
sources, with moderate BH masses ($\sim 10^{6-8} M_{\odot}$), which are 
hosted by spiral galaxies \cite{Komossa2008}. Since HE-emitting (radio-loud) 
NLSy~1 appear to be of the jetted AGN type, showing some blazar-like properties 
such as one-sided jets and superluminal motion, this has triggered discussion 
about the BH mass limit and accretion/merger history conducive for the origin 
and formation of relativistic jets. Relevant to this discussion is the question, 
whether jetted NLSy~1 might belong to a special subclass harbouring BHs 
with larger masses and hosted by elliptical galaxies instead. There are 
growing indications, however, that many of them are better modelled as 
disk-like galaxies \cite{Varglund2022}. For a discussion of these topics 
and related literature, the reader is referred to the recent overviews in 
refs.~\cite{DAmmando2019,Foschini2020}. 

\section{On Challenges and Progress in (jetted) AGN Physics}
\subsection{Challenges in AGN Physics}
\begin{figure}[htbp]
\begin{center}
\includegraphics[width = 0.98 \textwidth]{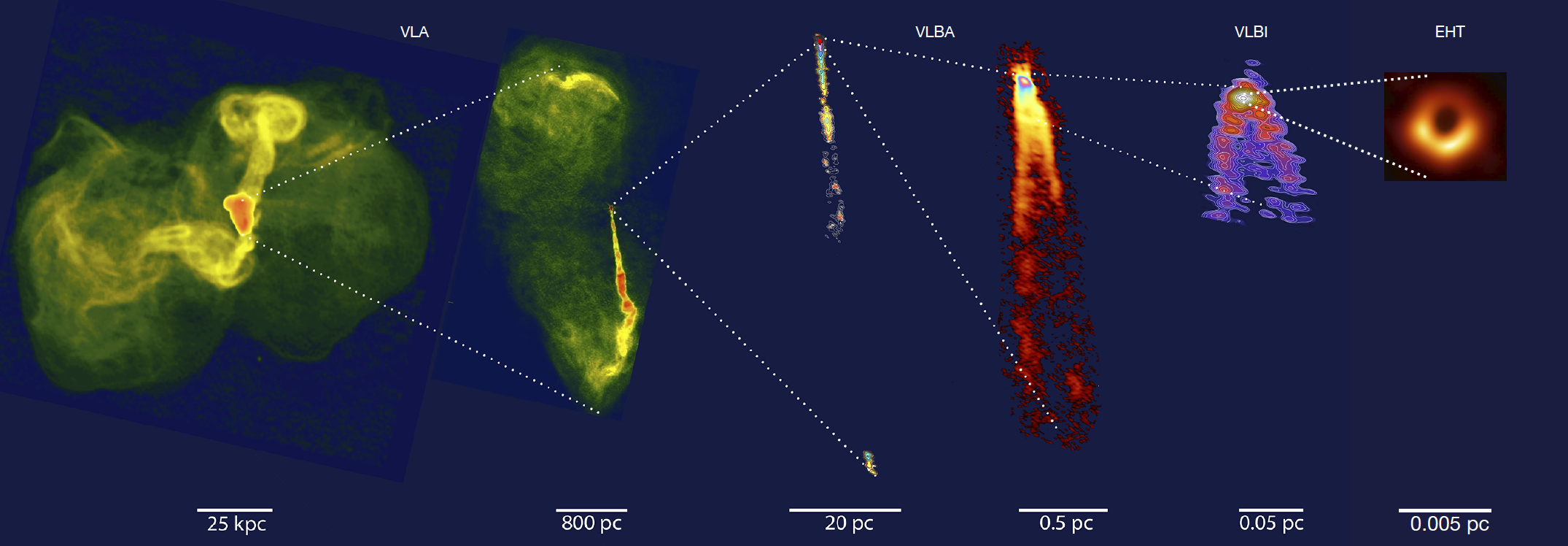}
\caption{The radio galaxy M87 as seen from large, radio-halo (VLA) down to 
black-hole (EHT) scales, corresponding to an "observed scale separation" 
of $\sim10^8$. Adapted based on ref.~\cite{Blandford2019}.}
\label{multi_scale}
\vspace*{-0.5cm}
\end{center}
\end{figure}

Understanding the physics of AGN jets and the role played by its supermassive 
BH is a challenging, multi-scale problem. Phenomenologically, the {\it observed 
scale separation} covers a range of almost ten orders (as in, e.g., Cen~A) of 
magnitude, see e.g. Fig.~\ref{multi_scale}.
This fact translates into a fundamental {\it physics and modelling challenge} of 
how to consistently bridge these different scales, e.g., of how to connect the 
global, source dynamical scale, the radiation (e.g., synchrotron, inverse Compton) 
scale and the dissipation (e.g., shock, reconnection, turbulence) scale.
While solid progress has been achieved over the years, no complete picture 
is existing up to now. In principle, this also applies to numerical simulations 
(cf.~\cite{Porth2019,Nishikawa2021} for related reviews).
While essential to inform and advance our understanding, they are not without
caveats, i.e., in general, AGN physics is also accompanied by a {\it methodological 
(computational) challenge}. In the context of jet formation for example,
conventionally employed {\it general-relativistic/magneto-hydrodynamic (GR/MHD)} 
simulations usually rely on an ambiguous numerical floor model; methodologically, 
it may seem rather surprising that a single fluid description is able to provide important 
insights (as it has) in cases where we expect the plasma to be collisionless; further, 
in ideal MHD there is little physical understanding of reconnection, particle acceleration 
and radiation, and so forth. On the other hand, first-principle {\it particle-in-cell (PIC)} 
simulations in astrophysics mostly deal with highly idealized setups, often in reduced 
dimensionality and for quite limited duration, along with artificially large gyro-radii. In 
fact, in the case of AGNs, the relevant scale separation (system size $r$ vs plasma 
skin depth $l_p$) can be as high as $r /l_p \sim 10^{6-8}$, cf. ref.~\cite{Ji2011}, and 
it remains unclear to which extent features seen in these simulations might persist up 
to the physical scale of interest. Notwithstanding these limitations, numerical simulations 
have become an indispensable tool for progress.

\subsection{Progress in AGN Physics}
The last years have seen both, significant progress and consolidation of knowledge 
in our understanding of the physics of AGNs. The following provides a short selection 
of exemplary results with relevance to gamma-ray emitting AGNs.\footnote{For 
recent results on hadronic processes and neutrino emission, the reader is referred 
to the related reviews by Paolo~Padovani and Elisa~Resconi.}

\subsubsection{Supermassive Black Holes in the Center of AGNs}
The presence of supermassive BHs in galactic nuclei has been suggested for more 
than half a century based on strong theoretical arguments. Early experimental efforts 
related to dynamical searches have been described $\sim25$ yr ago \cite{Kormendy1995}. 
Most recently, theoretical and experimental progress in BH research has eventually
been documented by the 2020 Nobel prize for Physics to Roger Penrose, Reinhard 
Genzel and Andrea Ghez. While the benchmark BH Sgr~A* is not (yet) an established 
gamma-ray emitter \cite{Katsoulakos2020}, the event horizon telescope (EHT) radio 
image of the BH shadow in M87 \cite{EHT2019} provides important information relevant 
to VHE research. 
The size $\simeq 2.5 r_s$ of the photon ring in M87 (where $r_s:=2 r_g := 2 GM_{\rm 
BH}/c^2$), implies a BH mass of $M_{\rm BH} \simeq 6.5 \times 10^9 M_{\odot}$ and 
constrains horizon scale (minimum) variability to $t_H\simeq 0.38$ d, close to what 
can be probed with VHE instruments \cite{RiegerL2018}. Further research along these 
lines (by, e.g., the next-generation EHT) will allow to probe deeper into the central engine 
in AGNs (BH - disk - jet, its connection and dynamics), and tighten the constraints on 
the current jet power (e.g., BH spin) relevant to VHE emission models.

\subsubsection{Convergence of theoretical, numerical \& observational evidence 
for jet stratification}
There is consolidation of knowledge that the jets in AGN are multi-layered, revealing
a lateral stratification similar to the one induced by a fast BH-driven (Blandford-Znajek: 
BZ, \cite{Blandford1977}) jet surrounded (and possibly confined) by a slower moving 
disk-driven outflow, cf.~\cite{Mizuno2022}.
Resolved (lateral) emission structures such as limb-brightening and linear polarisation 
signatures, e.g., refs.~\cite{Boccardi2017,Kim2018}, substantiate early two-flow and 
spine-sheath type non-thermal emission models \cite{Sol1989,Ghisellini2005}. In the 
case of M87, for example, significant structural patterns across the jet on sub-pc-scale 
have been detected, indicating the presence of both slow ($\sim 0.5c$) and fast 
($\sim0.92c$) components \cite{Mertens2016}. Pronounced edge-brightened features 
have now been seen in both, M87 and Cen~A \cite{Janssen2021}. In the high energy
context, characterising internal jet stratification is potentially important to, e.g., address 
the (putative) Doppler factor "crisis" in TeV (HBL) blazars \cite{Henri2006}, to develop 
more advanced acceleration (cf. Sec.~4.3) and emission models, and to probe into 
AGN unification scenarios (e.g., relative power of inner vs outer jet?).

\subsubsection{Acceleration and collimation of relativistic jets}
\begin{figure}[bhtp]
\begin{center}
\includegraphics[width = 0.6 \textwidth]{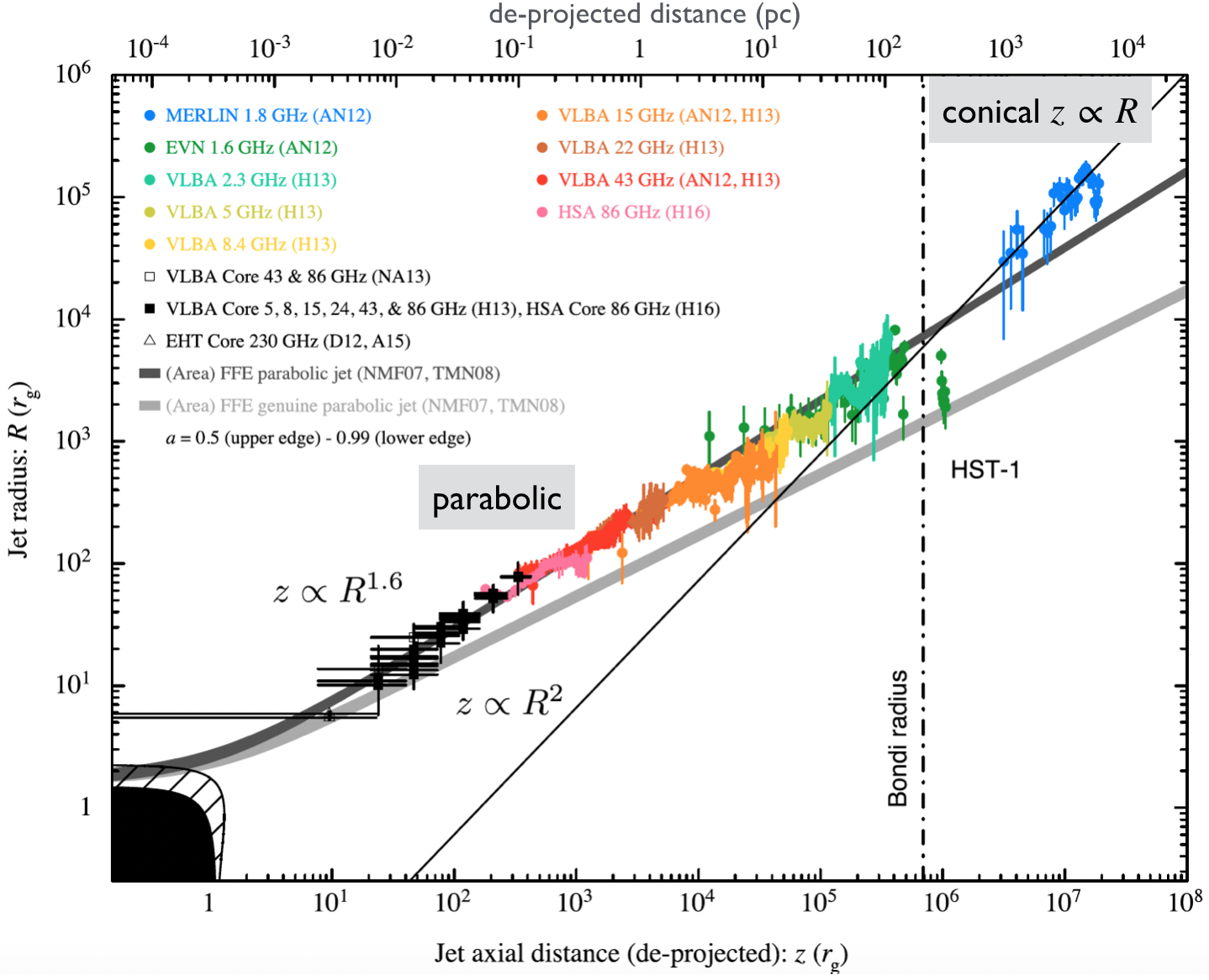}
\caption{The jet collimation profile for the radio galaxy M87 from sub-parsec to
kilo-parsec scales. From ref.~\cite{Nakamura2018}. \copyright AAS. Reproduced 
with permission.}
\label{m87_collimation}
\vspace*{-0.5cm}
\end{center}
\end{figure}

Detailed radio studies now provide insights into the jet collimation profile in more and
more nearby sources \cite{Hada2019,Kovalev2020}. In the case of M87 for example, 
the jet width profile is initially (semi-)parabolic and then transitions (at around the Bondi 
radius $r_B \sim 5 \times10^5 r_g \sim 150$ pc) to a conical shape, see 
Fig.~\ref{m87_collimation}. There is evidence that the radio flow is initially slow (on 
scales $\sim 0.03$ pc) and gradually accelerates with distance, reaching $\Gamma\beta 
\sim 3$ on scales of $r_B$ \cite{Nakamura2018,Park2019}. 
This can be understood in terms of a jet that is magnetically dominated at its base and 
whose magnetic energy is gradually converted into kinetic energy while experiencing 
some external confinement \cite{Beskin2006,Komissarov2007}. These results are 
instructive for improving our understanding of the non-thermal emission in low-luminosity 
FR~I type sources, and of the dynamical role (confinement) played by a disk-driven wind. 
While consequences of flow deceleration on high-energy SED modelling have been 
explored in the past \cite{Georganopoulos2005}, consideration of flow acceleration 
seems an important desideratum.

\subsubsection{Convergent evidence for relativistic motion in large-scale AGN jets}
While jets experience deceleration on larger scales, they can still be substantially 
relativistic on kilo-parsec scales, even for moderately powerful sources. The nearby 
($d\sim 95$ Mpc) FR~I radio galaxy 3C~264 is a prototype example (with estimated 
jet power $L_j \sim 5 \times 10^{43}$ erg/s) \cite{Meyer2011}, exhibiting superluminal 
motion and collision of knots (internal shocks) in HST observations on scales of 
$\sim (0.5-1)$ kpc, corresponding to a bulk flow Lorentz factor $\Gamma \sim 7$
and a Doppler factor $D\sim 7$ for an inclination $\theta\sim 8^{\circ}$ \cite{Meyer2015}. 
These findings are of interest not only because 3C264 is a recently detected VHE 
emitter \cite{Archer2020}, but also by offering general insights into particle acceleration 
at shocks, and the role of nearby large-scale jets as putative sites of ultra-high energy
cosmic ray (UHECR) acceleration.

\subsubsection{Extreme TeV blazars}
A sizeable fraction of VHE-emitting AGNs ($\sim1/4$ of all HBL) are now composed 
of BL Lac objects with hard intrinsic VHE spectra, i.e. with de-absorbed power-law (PL)
photon indices $\Gamma_{\rm VHE} \lesssim 1.5-1.9$, implying an SED bump peaking 
above $1$ TeV, see ref.~\cite{Costamante2020} for a review. The prototypical source of 
these extreme highly peaked BL Lacs (EHBLs) is 1ES 0229+200 ($z=0.14$) with an SED peak 
(de-absorbed) above $4$ TeV. The SED modelling of these sources is challenging, and 
in the common synchrotron-self Compton (SSC) approach requires unusual (if not 
extreme) conditions such as an electron distribution with a very high minimum Lorentz 
factor $\gamma_{\rm min} \gtrsim 10^{4}$, or one following a relativistic Maxwellian 
\cite{Lefa2011}. The narrow pile-up feature seen in Mkn~501 \cite{MAGIC_MKN501}, 
cf. Fig.~1 (left), might be reminiscent of the latter \cite{Lefa2011b}. 
These extreme TeV blazars are particularly interesting as they indicate emitting 
regions significantly away from equipartition ($u_B/u_e \lesssim 10^{-4}$ in SSC). 
The implied low magnetisation $\sigma$ would seem to disfavour reconnection and 
instead point to stochastic or multiple shocks as possible acceleration mechanism (see 
also Sec.~4.2). VHE variability studies will enable to probe into this as only modest 
($\geq$ month-type) intrinsic variability would be expected (unless geometrical effects 
would interfere). Unusual spectral shapes at VHE energies will complicate EBL 
(Extragalactic background light) studies using simple extrapolation functions (for a
recent review of developments, see e.g. ref.~\cite{Biteau2022}).

\subsubsection{Gamma-Ray Astrophysics in the Time Domain}
Within the last decade, gamma-ray astronomy has successfully entered the time domain 
\cite{Rieger2019_time}. Key examples include the detection and characterisation of 
ultra-fast VHE variability (down to a few minutes) \cite{Aharonian2017,Costamante2020} 
as well as evidence for year-type quasi-periodic oscillations (QPOs) in the Fermi-LAT light 
curves of blazars \cite{Ait_Benkhali2020,Penil2022}. A prominent example of the latter is 
the HBL object PG 1553+113 ($z\sim 0.5$) with an (observed) HE period of $\simeq 2.2$ yr, 
possibly caused by a close supermassive binary BH system (SBBHs) \cite{Sobacchi2017,
Tavani2018}. Application of more advanced time series techniques, exploring power-law 
noise characteristics and log-normality has been steadily growing \cite{HESS2017_PKS}. 
Theoretical efforts are now gaining momentum, cf. \cite{Rieger2019_time} for a review, 
and refs.~\cite{Polkas2021,Dmytriiev2021} for selected contributions. Properly 
understanding the variability characteristics has important implications for the origin of 
the emission (e.g., location and driving process) and the overall source dynamics and 
evolution (e.g., SBBHs, disk physics).

\section{Selected Theoretical Advances}
The following subsections aim to provide exemplary and more detailed insights into 
three different fields where theoretical advances in AGN physics, whether conceptual 
or methodological, have been significantly influenced by gamma-ray observations.
 
\subsection{Magnetospheric Processes in AGNs}
\begin{figure}[htbp]
\begin{center}
\includegraphics[width = 0.75 \textwidth]{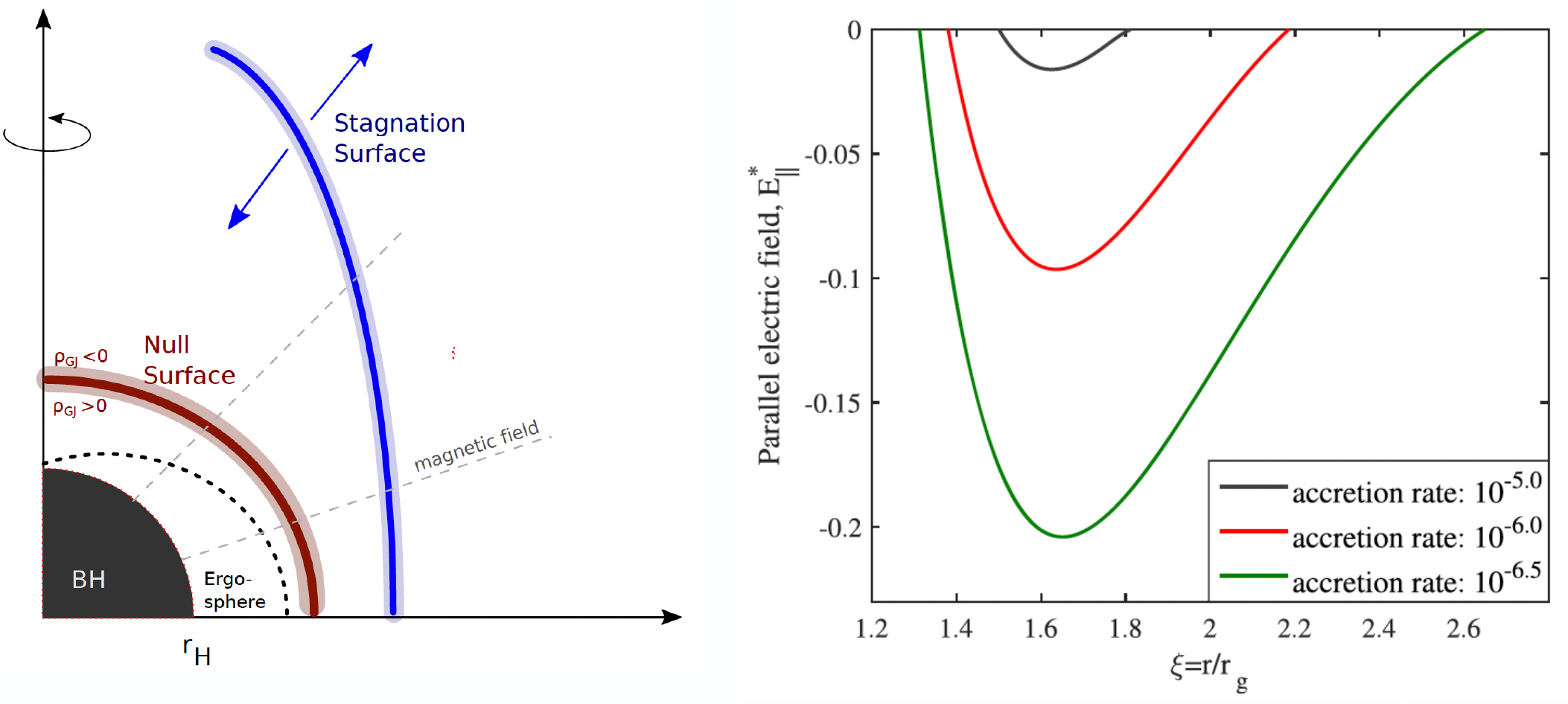}
\caption{{\it Left:} Illustration of possible locations of charge-deficient regions (gaps) 
around a rotating BH. The red line denotes the null surface across which $\rho_{\rm 
GJ}$ changes sign, while the blue line delineates the stagnation surface from which 
stationary MHD flows start. {\it Right:}
Exemplary gap evolution for a Kerr BH of $10^9 M_{\odot}$ accreting in a radiatively 
inefficient mode. The curves show the radial distribution of the parallel electric field 
around the null surface for different accretion rates $\dot{m}$. The gap width and 
voltage drop increase as $\dot{m}$ drops because pair creation of energetic photons 
with the soft photon field becomes less efficient. From ref.~\cite{Katsoulakos2020a}.}
\label{gap}
\vspace*{-0.4cm}
\end{center}
\end{figure}

\begin{figure}[htbp]
\begin{center}
\includegraphics[width = 0.68 \textwidth]{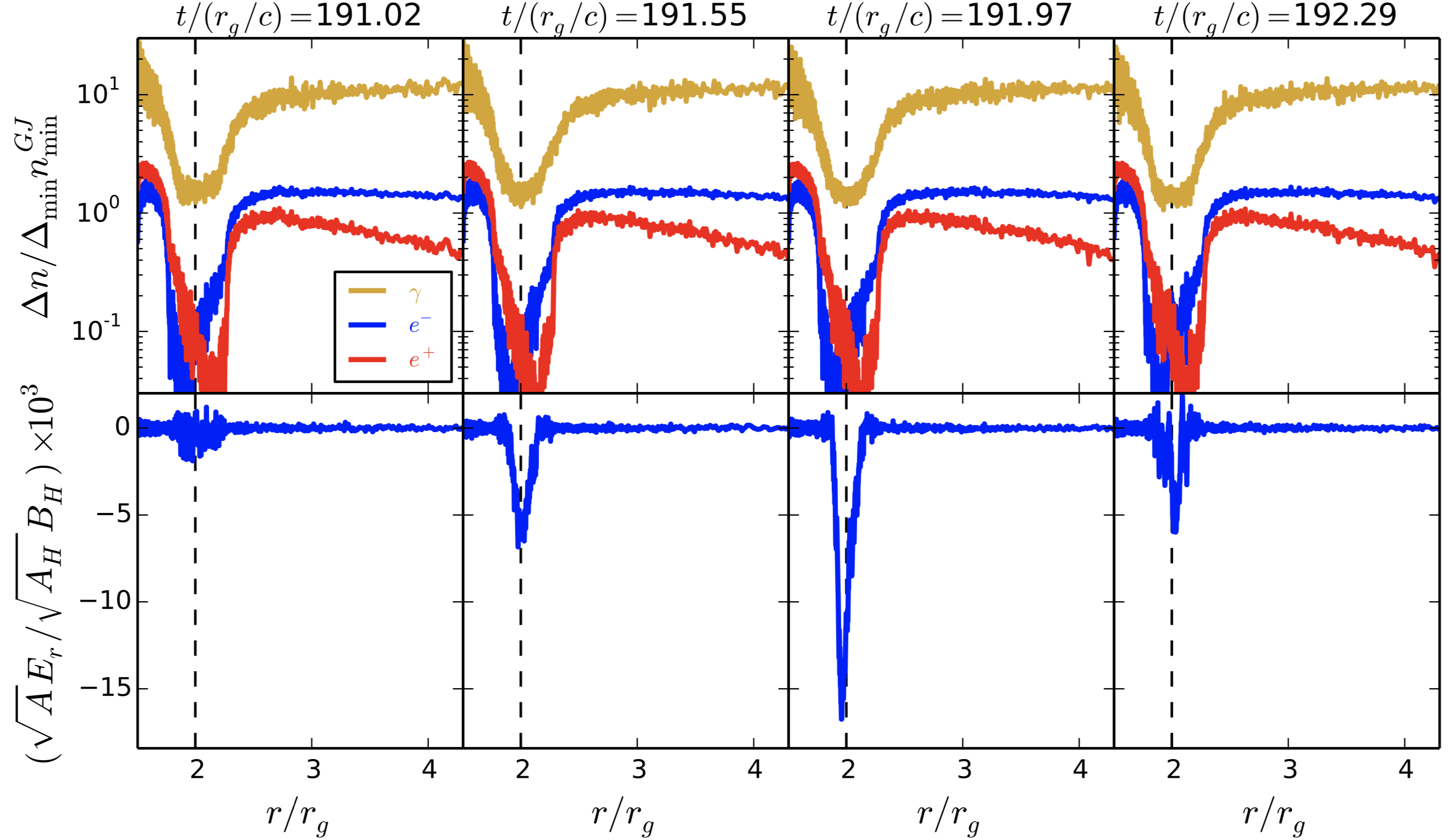}
\caption{Exemplary gap cycle seen in 1d PIC simulations (spit monopole field) 
for a steep ambient photon field ($n_s$) with $\epsilon_{\min}=0.0005$ eV and 
fiducial Thomson depth $\tau_0:= n_s\sigma_{\rm Th} r_g=100$ (yielding a pair 
creation length $\sim0.3 r_g$). Top panel: Evolution of the normalized 
densities of electrons, positrons and photons. Bottom panel: Evolution of 
the electric field. Vertical dashed line denotes the null surface. From 
ref.~\cite{Kisaka2020}. \copyright AAS. Reproduced with permission.}
\label{Kisaka_gap}
\vspace*{-0.5cm}
\end{center}
\end{figure}

The detection of VHE variability on BH horizon crossing time scales ($\sim r_g/c$), as 
in e.g. M87, has been a significant driving force for the development of magnetospheric 
particle acceleration and emission models, see e.g. ref.~\cite{RiegerL2018} for a review.
Magnetic fields that are brought in and dragged into rotation ($\Omega_F$) by the BH
can induce an electric field component (parallel to the magnetic field), that, if not 
screened, can facilitate efficient particle acceleration. This could potentially occur either 
around the so-called null surface across which the Goldreich-Julian \cite{Goldreich1969}
charge density, $\rho_{\rm GJ}\sim(\Omega_F- \omega) B$, required to screen the electric 
field, changes sign, or at the stagnation surface that separates MHD in- and outflows, see 
Figure~\ref{gap}.
The maximum available voltage drop for a BH gap of width $h$ is of the order of 
\cite{Katsoulakos2018} 
\begin{equation}\label{voltage_drop}
\Delta \Phi = \frac{1}{c} \Omega_F r_H^2 B_H \, (h/r_g)^2  
                     \simeq 2\times 10^{21} \dot{m}^{1/2} M_9^{1/2}\,(h/r_g)^2 \;\mathrm[V]\,,
\end{equation} for a horizon-threading field $B_H \simeq 2\times 10^5 \dot{m}^{1/2} 
M_9^{-1/2}$ G, where $M_9 \equiv M_{\rm BH}/10^9 M_{\odot}$, $\Omega_F= 
\Omega_H/2=c/4r_g$, and $\dot{m} \equiv \dot{M}/\dot{M}_{\rm Edd}$.  
Electrons that get accelerated in these gaps produce $\gamma$-rays via curvature 
radiation and inverse Compton (IC) up-scattering of ambient soft photons, producing
an observable and variable VHE signal \cite{Levinson2011}. 
Interaction ($\gamma\gamma$-absorption) of energetic $\gamma$-rays with 
low-energy soft photons can trigger a pair cascade that generates plasma and 
ensures closure of the gap \cite{RiegerL2018}. 
On conceptual grounds, magnetospheric gaps thus provide a self-consistent means 
for the continuous plasma supply ("numerical floor") needed to activate a force-free 
(BZ) MHD outflow \cite{Contopoulos2018}. 
Correspondingly, unless accretion rates are sufficiently low (i.e., well below $\dot{m} 
\sim 10^{-4}$), efficient pair production will occur (if non-negligible gaps form at all, 
cf.~\cite{Levinson2011,Hirotani2016}), leading to gap sizes $h \ll r_g$, thereby 
significantly reducing the accessible gap potential, cf. eq.~[\ref{voltage_drop}]. 
Typically, achievable (radiation-limited) electron Lorentz factors are constrained to 
$\gamma_e \sim 10^{8-10}$ \cite{Katsoulakos2018}. To improve insights into 
characteristic electric field strengths and achievable voltage drops, one can seek 
for self-consistent steady gap solutions \cite{Hirotani2016,Katsoulakos2020a} by 
solving the system of relevant partial differential equations (e.g., Gauss' law, 
equation of particle motion, continuity equation), see e.g. Figure~\ref{gap} (right).
For example, taking an accretion rate $\dot{m}=10^{-5.75}$, which is close to the 
mean MAD value used in EHT-GRMHD simulations for M87 \cite{EHT2019b} 
and which corresponds to jet powers of a few times $10^{43}$ erg/s, the gap 
properties for M87 (assuming the inner disk to be ADAF-type) can be evaluated 
\cite{Katsoulakos2020a}. The resultant gap sizes are of the order of $\sim0.8 r_g$, 
suggesting that the VHE emission in M87 might be variable down to timescales of 
$\sim0.4$ days. The inferred gap power of $\sim 5 \times10^{41}$ erg/s would 
make it in principle possible to accommodate the VHE emission seen during its 
high states \cite{Ait_Benkhali2019}. The achievable voltage drop would be of the 
order of $\sim 10^{18}$ V, suggesting that the BH in M87 is not an efficient UHECR 
proton accelerator. The appreciation that gap operation is expected to be intermittent, 
has in recent times triggered a variety of time-dependent investigations using PIC 
simulations, see e.g. refs. \cite{Chen2020,Crinquand2020,Kisaka2020,Kisaka2022}. 
Relative comparison is often not straightforward as different setups and complexities 
have been employed (e.g., with regard to box sizes, cell numbers, run times, 1d/2d, 
radiation reaction, choice of soft photon field). Generically, the outcome is highly 
sensitive to the assumed ambient soft photon field (i.e., minimum photon energy 
$\epsilon_{\min}$ and spectral shape, e.g., PL index), making an adequate description 
an important prerequisite as it determines the efficiency of pair cascades. There are 
indications for a periodic opening (timescale $\sim r_g/c$) of macroscopic gaps (with 
$h \sim [0.1-1]\, r_g$), cf. Fig.~\ref{Kisaka_gap}. 
Issues that deserve further studies concern e.g., the degree to which detected
oscillation periods could be dependent on the simulation setup (e.g., box size), 
and the extent to which the emerging particle distributions might be 
bimodal. Insights gained will help to better characterise the link between gap 
activity, VHE flaring and jet formation.

\subsection{Modelling Parsec-Scale Jet Emission in Blazars}
The spectral energy distributions (SEDs) of gamma-ray blazars have for long 
been modelled by leptonic (SSC) and/or hadronic (e.g., $p\gamma$) radiation
processes, see e.g. \cite{Cerruti2020,Sol2022} for a review. Quite often their 
SEDs can be satisfactorily reproduced with different sets of assumptions, which 
significantly limits inferences on the source physics. Incorporating microphysical 
insights from kinetic plasma simulations into SED modelling can allow an important 
step forward. 
One exemplary context relates to the requirement of high minimum electron Lorentz 
factor $\gamma_{\rm min,e}$ and low magnetisation $\sigma$ in SSC models for 
extreme TeV blazars (cf. Sec.~3.2.5). A recently proposed framework considers 
co-acceleration (diffusive shock) of electrons and protons at mildly relativistic
(e.g., $\Gamma_s \sim 3$), weakly magnetised (internal or recollimation) shocks 
in blazar jets \cite{Zech2021}.
Shocks are known to convert a significant fraction of the kinetic energy of the 
incoming plasma flow to thermal energy, represented by a Maxwellian particle 
distribution $n(\gamma) \propto \gamma^2 \exp[-\gamma/\Theta]$ with $\Theta
 \equiv k_B T/mc^2$. Assuming an efficient energy transfer (heating/thermal 
coupling) from protons to electrons in the shock transition layer, one can write
$k_B T_P \sim \Gamma_s m_p c^2$ and $T_e = \xi T_p$ where $\xi < 1$, with 
$\Gamma_s$ the Lorentz factor of the shock. If the peak in the Maxwellian 
electron distribution is identified with $\gamma_{\rm min,e}$, then roughly 
$\gamma_{\rm min,e} \sim  k_B T_e /(m_e c^2)$, i.e.
\begin{equation}
\gamma_{\rm min,e} \sim \left(\frac{m_p}{m_e}\right) \Gamma_s \xi \simeq 
1800\,\Gamma_s \xi
\end{equation}
Hence, given suitable conditions, a high $\gamma_{\rm min,e}$ is naturally 
expected. In PIC simulations of highly relativistic shocks, efficient Fermi-type 
particle acceleration has only been seen for weakly magnetized (e.g., $\sigma  
\lppr 10^{-3}$) shocks, in line with considerations in the above framework. 
Since these shocks will also generate magnetic (micro)turbulence, the effective 
magnetization ($\sigma := B^2/(4\pi \langle \gamma \rangle n m c^2$) could 
in principle be higher than the pre-shock one, yet decaying downstream with 
distance. Accordingly, the magnetization in the radiation zone, $\sigma_{\rm 
rad}$, may be in the range $\sigma \lppr \sigma_{\rm rad} \lppr 0.01$.\\ 
For high (Alfvenic) Mach numbers, significant thermal coupling can occur. 
\begin{figure}[htbp]
\begin{center}
\includegraphics[width = 0.95 \textwidth]{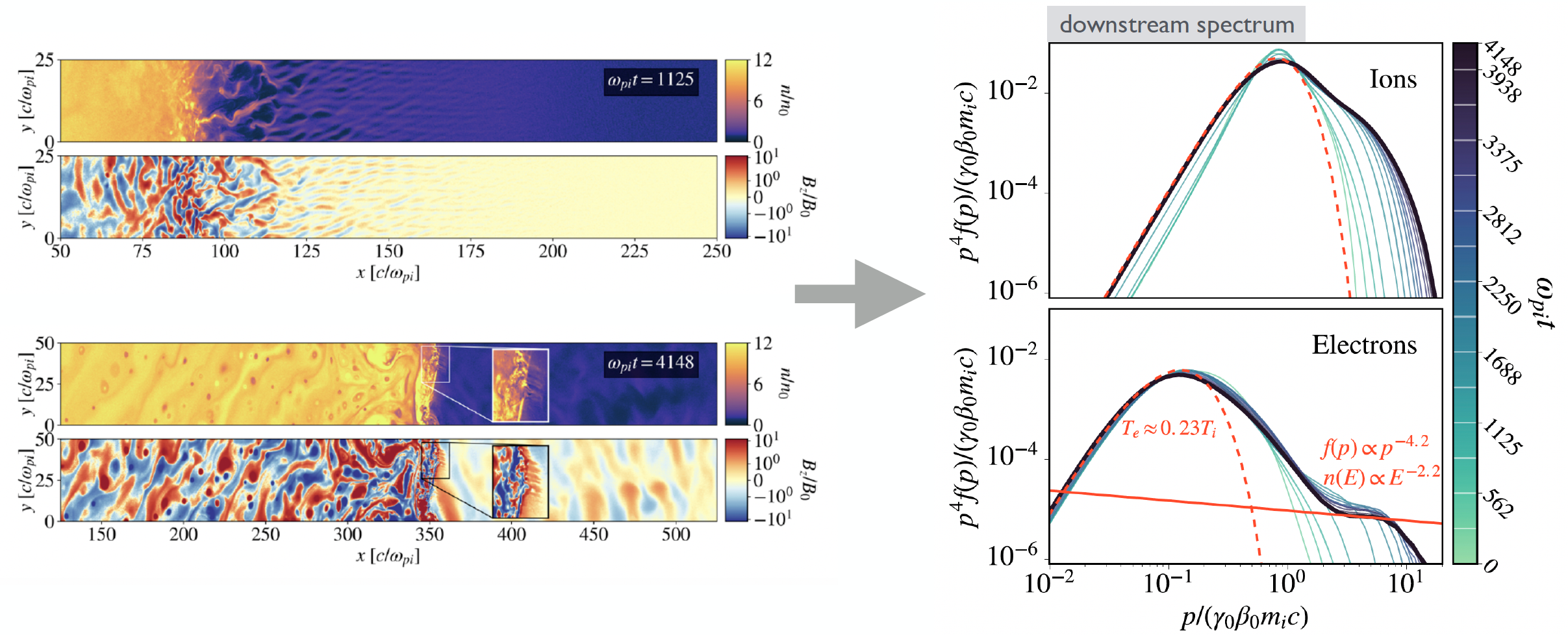}
\caption{2d PIC simulations of a weakly magnetized, mildly 
relativistic, high Mach number ($M_A=15$) shock with $\Gamma_s\simeq1.7$.
{\it Right:} Shock evolution and magnetic field generation (initially Weibel-, later 
Bell-mediated) with time. {\it Right:} Downstream particle distributions $p^4 f(p)$ at 
different times, with the formation of a non-thermal power-law tail on top of a 
thermalized particle distribution. At late times, the downstream electron temperature 
is $20-30\%$ of the ion temperature, indicating significant coupling. 
Based on ref.~\cite{Crumley2019}.}
\label{PIC_heating}
\end{center}
\vspace*{-0.5cm}
\end{figure}
In large 2d PIC simulations of quasi-parallel, mildly relativistic ($\beta_s\simeq 0.8$), 
weakly magnetized ($\sigma=0.007$, defined downstream) electron-ion shocks (with 
reduced mass ration $m_p/m_e=64$), for example, $T_e \sim (0.2-0.3)\, T_p$ and 
$k_B T_p \sim 0.2 m_p c^2$ has been seen, cf. Fig.~\ref{PIC_heating} \cite{Crumley2019}. 
For smaller magnetizations and larger shock speeds, electron preheating can be further 
increased \cite{Sironi2013,Vanthie2020}. Figure~\ref{Zech_Lemoine} shows an 
exemplary reproduction of the SED of the prototypical EHBL 1ES~0229+200 
motivated by such a framework. 
\begin{figure}[htbp]
\begin{center}
\includegraphics[width = 0.82 \textwidth]{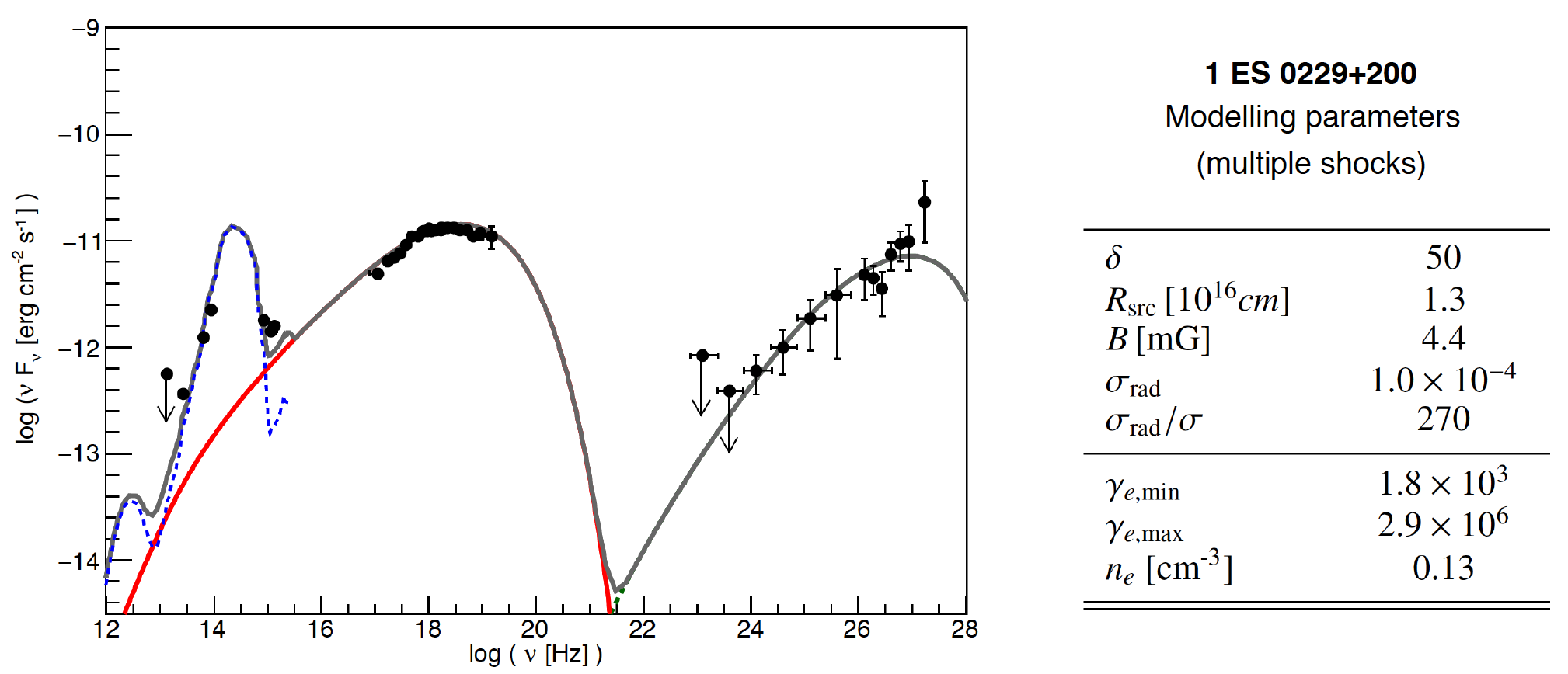}
\caption{Reproduction of the SED of the EHBL source 1ES~0229+200 
in a multiple shock scenario with input parameters for the initial electron distribution 
as shown in the table to the right. From ref.~\cite{Zech2021}, reproduced with 
permission~\copyright~ESO.}
\label{Zech_Lemoine}
\vspace*{-0.5cm}
\end{center}
\end{figure}
In this case, re-acceleration at multiple ($n \geq 2$) shocks is needed to account 
for the hard particle spectra required for spectral fitting. This leads to a modification 
(hardening and power-law deviation) of the input, single shock electron spectrum 
(of power law momentum index $s=2.2$), and also increases the effective injection
Lorentz factor to $\gamma_{\rm min, eff} \sim \Gamma_s^{2n/3}\gamma_{\rm min, 
e}$ \cite{Zech2021}. The noted framework represents an 
exemplary approach to utilise microphysical insights for blazar SED modelling.
One central assumption is that the emission arises in a jet region that is kinetically 
dominated by an electron-proton composition (with protonic emission expected to 
be suppressed) rather than an $e^+e^-$ one. Similar indications have been found
in the case of another prominent $\gamma$-ray blazar, PKS 2155-304, based on 
a statistical analysis of its X-ray emission \cite{Jagan2021}, though there is also 
other evidence (based on optical circular polarization) suggesting that the (non-thermal)
positron fraction cannot be too low \cite{Liodakis2022}, cf. also ref.~\cite{Romero2021}. 
A possible caveat for the noted approach relates to the model requirement of a 
region significantly out of equipartition and characterized by a very low 
magnetic field. In principle, variability studies could allow to probe into these 
assumptions.\\ 
A related framework considers the electrons in the jet to be (initially) diffusively 
accelerated at a recollimation-type shock, but then further energized through 
stochastic (2nd order Fermi) acceleration in the downstream flow \cite{Tavecchio2022}. 
Hence, instead of acceleration at multiple shocks, recollimation in a weakly 
magnetized flow is presumed to result in strong turbulence. While the synchrotron 
spectrum can be satisfactorily reproduced in a SSC model where synchrotron 
cooling is comparable to particle escape (cf. also \cite{Lefa2011} for another 
variant), the Compton SED component appears somewhat too hard. 
Since EHBL SED models are in general far from equipartition (i.e., $\sigma \ll1$), 
characteristic Alfven speeds are sub-relativistic ($v_A \propto \sqrt{\sigma} \ll c$) 
making stochastic acceleration less efficient ($t_{\rm} \propto 1/v_A^2$), so that 
it remains to be explored to what extent a self-consistent SED fit can be achieved.
Nevertheless, these approaches represent instructive examples of how to fruitfully
combine insights from jet simulation, particle acceleration physics and SED 
modelling.

\subsection{Understanding the large-scale jet emission of AGNs}
\begin{figure}[htbp]
\begin{center}
\includegraphics[width = 0.98 \textwidth]{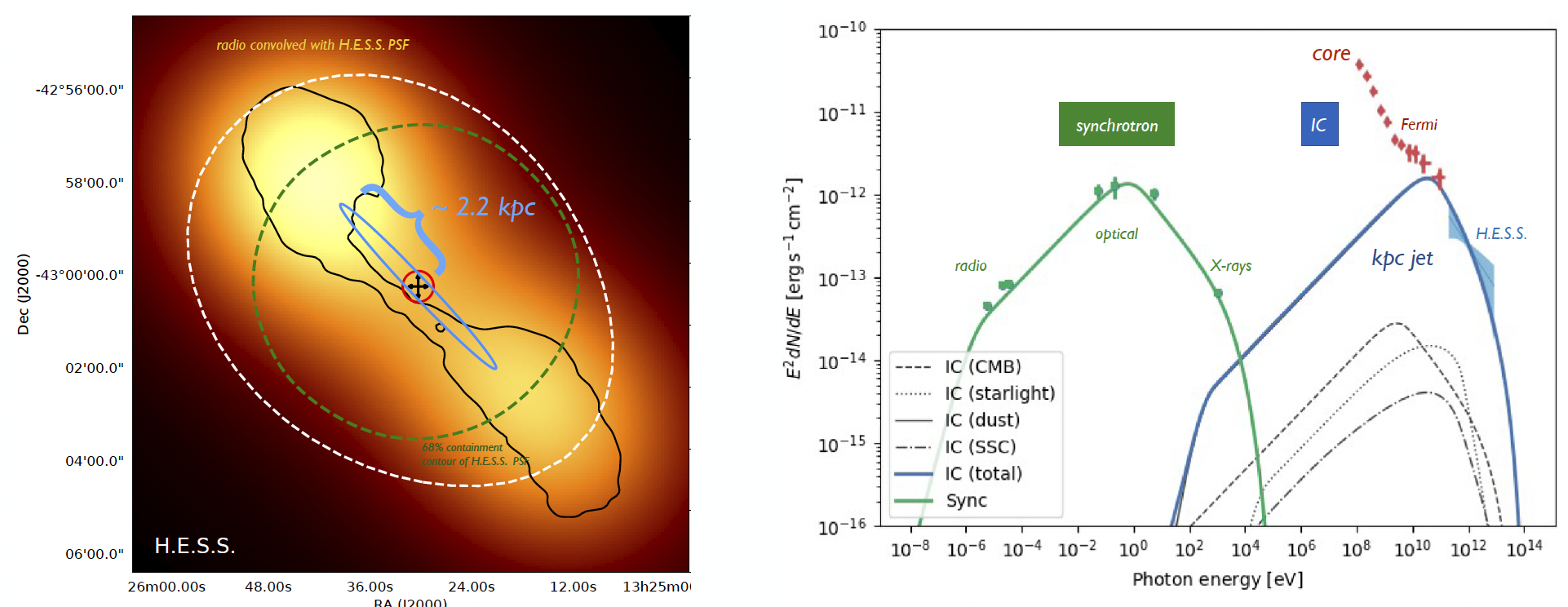}
\caption{Extended VHE emission along the kpc-scale jet of Cen~A as seen by 
H.E.S.S. The SED of the kpc-scale jet can be reproduced by means of an 
(exponential-cutoff) broken-power law electron distribution with $\gamma_b
\simeq 10^6$ and cut-off energy $\gamma_c\simeq 10^8$, for a characteristic
jet magnetic field strength of $B\simeq 20 \mu$G. 
Interestingly, IC emission by the kpc-scale jet may be behind the unusual 
spectral hardening \cite{HESS_FERMI_CENA} seen at GeV energies. 
Based on ref.~\cite{HESS_CenA}.}
\label{HESS_CenA}
\vspace*{-0.5cm}
\end{center}
\end{figure}

The origin of extended X-ray emission along the large-scale jets in AGNs has for 
long been a matter of debate, with electron-synchrotron and IC-CMB scenarios 
as the leading contenders, see e.g. \cite{Harris2006,Georg2016} for review. 
Within more recent years, however, IC-CMB models have been disfavoured in 
an increasing number of sources as they tend to, e.g., over-predict Fermi-LAT 
gamma-ray flux limits \cite{Breiding2017,Breiding2023} or are not supported by
detailed X-ray spectral information \cite{Sun2018}. 
An electron synchrotron origin, however, requires to maintain electrons with 
Lorentz factors $\gamma_e\sim10^8$ all along the jet. The short cooling 
timescales $t_{\rm cool}  \propto 1/\gamma$ of these electrons, and 
correspondingly short cooling lengths $c t_{\rm cool} \ll 1$ kpc, calls for the 
operation of a "distributed" acceleration mechanism along the jet, beyond 
"localised" shock acceleration. The recent detection of extended VHE emission 
along the kpc-scale jet of Cen~A  \cite{HESS_CenA} has provided an important 
validation to this picture, see Fig.~\ref{HESS_CenA}, by directly tracing the 
electrons and removing the degeneracy ($B,\gamma$) inherent in an X-ray 
synchrotron model. IC up-scattering of dust by ultra-relativistic (synchrotron
X-ray emitting) electrons with $\gamma_e =10^8$ accounts for the observed 
VHE emission and confirms the X-ray synchrotron interpretation. The results
imply a jet that is only weakly magnetized ($\sigma <0.5$) on kiloparsec 
scales.
Stochastic (2nd order Fermi) and shear particle acceleration processes are among 
the most promising mechanisms facilitating continuous re-acceleration in kinetically 
dominated jets \cite{Liu2017,Tavecchio2021,Wang2021,Wang2022}.
Particle acceleration in the latter framework essentially draws on the jet flow velocity 
difference that a charged particles experiences as it moves across the jet shear (cf.
Sec.~3.2.2), see e.g. ref.~\cite{Rieger2019} for an accessible introduction. 
It can be viewed as a second-order ($\Delta E \propto u_{\rm sc}^2 E$) Fermi-type 
acceleration process, in which the common scattering center speed, $u_{\rm sc}$, 
is replaced by an effective velocity, $\bar{u}$, determined by the shear flow profile, 
i.e., by the characteristic flow velocity change sampled over the mean free path 
$\lambda\simeq \tau c$ of the particle. Hence, for instance $\bar{u} = (\partial u_z/
\partial r)\, \lambda$ for a continuous shear profile $\vec{u} =  u_z(r) \vec{e}_z$. 
This results in a characteristic acceleration timescale
\begin{equation} 
t_{\rm acc} \simeq \frac{E}{dE/dt} \simeq \frac{\tau}{\left<\Delta E/E \right>}
 \propto \frac{\tau}{\bar{u}^2} \propto \frac{1}{\lambda}\,,
\end{equation} scaling inversely with the particle mean free path $\lambda$ or 
scattering time $\tau$. While this makes shear flows also interesting for UHE 
cosmic-ray acceleration in the jets of AGNs \cite{Rieger2022CR}, it typically 
requires some energetic seed injection for electron acceleration to proceed 
efficiently. In principle, this could be achieved by shock or classical 2nd order 
Fermi acceleration (for which $t_{\rm acc} \propto \lambda$) \cite{Liu2017}, or 
possibly, by kinetic-scale reconnection processes \cite{Sironi2021}. 
Phenomenologically,  this could result in multi-component particle distributions, 
e.g., broken power-laws with changes in spectral indices different from a 
classical cooling break. A recent application shows that in the presence of a 
Kolmogorov-type $\lambda \propto \gamma^{1/3}$, the SED of the kpc-scale 
jet in Cen~A can be successfully reproduced within a linear shear model for 
spine velocities of $v_j \simeq 0.7$c and shear layer widths of $\Delta r \simeq 
100$ pc \cite{Wang2021}. In general, efficient shear acceleration requires 
relativistic flow speeds \cite{Webb2018,Rieger2022} and this confines its
applicability to astrophysical objects with fast outflows. There is currently a
promising trend to incorporate particle acceleration and non-thermal emission 
recipes in large-scale MHD jet simulations, e.g. \cite{Mukherjee2021}. While 
such a hybrid approach is vital, it should be kept in mind that it still involves 
significant extrapolations (cf. Sec.~3.1.) that may need to be further explored. 
The prospects are good, though, for achieving further progress soon as 
several new observational facilities are coming online \cite{Blandford2019}.  

\section{Conclusions}
By providing new and often unexpected insights, multi-wavelengh and 
multi-messenger observations have become an indispensable tool for progress 
in many areas of astronomy. 
The field of AGNs, and in particular jetted AGNs, to which most extragalactic 
gamma-ray sources belong, is no exception in this regard \cite{Boettcher2019}. 
In the present review, I have highlighted some of the developments that are of 
particular relevance for our understanding of the nature of these sources, 
and/or directly driven and inspired by gamma-ray observations, ranging from 
new observational findings to conceptual and methodological advances. I 
believe that we have seen true progress in our understanding of the physics 
of gamma-ray loud AGNs. As for jetted AGNs in general and for gamma-ray 
loud AGNs in particular: "The observational prospects for securing our 
understanding of AGN jets are bright." (Blandford et al., \cite{Blandford2019}). 
In particular, given available (from e.g. Fermi-LAT and the current IACTs, 
HESS, MAGIC and VERITAS to HAWC, LHAASO) and upcoming capabilities 
(CTA) at highest photon energies.

\acknowledgments {\it I would like to thank Josep Maria Parades, Valenti 
Bosch-Ramon, Pol Bordas, Marc Ribo and the whole Barcelona LOC team 
for a truly pleasant and inspiring conference. I am grateful to Felix Aharonian 
for fruitful collaboration over the years. Funding by a DFG Fellowship 
under RI~1187/8-1 is gratefully acknowledged.}

\bibliographystyle{JHEP}
\bibliography{references}

\end{document}